\documentclass[twoside,letterpaper]{article}
\usepackage{verbatim}
\usepackage{moreverb}
\usepackage{url}
\usepackage{amsmath}
\usepackage{color}
\usepackage{appendix}
\usepackage{graphicx}
\usepackage{subfig}
\usepackage[colorlinks=true,bookmarks=true,pdfauthor={Vadim Zaliva, lord@crocodile.org},
            pdftitle={Platform-Independent Firewall Policy Representation},
            pdftex]{hyperref}
\usepackage{listings}
\usepackage{setspace} 
\author{Vadim Zaliva, lord@crocodile.org}
\date{\today}
\title{Platform-Independent Firewall Policy Representation}

\begin{document}
\lstset{language=XML,basicstyle=\tiny,markfirstintag=true,breaklines=true}

\maketitle

\begin{abstract}
  In this paper we will discuss the design of abstract firewall model
  along with platform-independent policy definition language. We will
  also discuss the main design challenges and solutions to these
  challenges, as well as examine several differences in policy
  semantics between vendors and how it could be mapped to our
  platform-independent language. We will also touch upon a processing
  model, describing the mechanism by which an abstract policy could be
  compiled into a concrete firewall policy syntax. We will discuss
  briefly some future research directions, such as policy optimization
  and validation.  \emph{Keywords:} firewall, policy, NAT, fwbuilder,
  security, rules

\end{abstract}

\pagebreak 
\tableofcontents

\pagebreak 
\section{Introduction}

Presently, firewall administrators are often required to manage
multiple firewall platforms from different vendors. Each of these
platforms has its own language to describe firewall policies. Besides
syntax differences, firewall policy models also vary from vendor to
vendor. If we make a parallel to programming languages, a firewall
administrator is required to learn multiple assembly languages. One
possible solution is the introduction of a high-level,
platform-independent firewall policy description language which could
be compiled into representations specific to particular platforms.
This approach relieves the burden on firewall administrator of
learning the low-level details of multiple firewall
platforms. Additionally, it helps to eliminate large groups of trivial
errors which a human could make during policy configuration, by
allowing a user to work with higher level abstractions without being
burdened by low-level policy syntax details. Having a
platform-independent policy representation will also allow the user to
develop a class of cross-platform tools for managing, analyzing, and
validating such policies. We believe that our approach will allow
administrators to increase system security by reducing the chance of
human error.

The ideas described in this paper are implemented in a successful open
source project called {\sl Firewall Builder}\cite{kurland2003fb}. It
currently supports five firewall platforms and is included in major
Linux distributions. Firewall Builder allows the user to create and
edit policies of an abstract firewall expressed in a
platform-independent language. The project provides convenient GUI for
editing firewall policies. The abstract policy uses a set of provided
{\sl policy compilers} to compile into policy files for concrete
firewall platforms. In this paper we have focused on abstract firewall
models and policy compilation. We refer readers to related documents
on Firewall Builder user interface\cite{fbug2003}, API,
extensibility\cite{fwbupgrades}, etc.

This paper is organized as follows: In section~\ref{abstractfirewall}
we wil describe the Abstract Firewall model we are using. In
section~\ref{challanges} we will discuss some examples of
platofrm-specific differences to illustrate the kinds of problems we
are solving. Then, in section~\ref{processingtechniques}, we will
discuss some processing techniques we have used. Finally, in
section~\ref{future} we will cover some of the possible directions of
future research.

\section{Abstract Firewall}
\label{abstractfirewall}

Firewall Builder presents a user with a Synthetic Model of a firewall,
in which we can combine features supported by various firewall
platforms. We also made some assumtions about the semantics of some
rules, which are normally also platform dependent.

When working with Firewall Builder, the user only needs to know this
abstract firewall model. The user defines policy for this imagninary
abstract firewall, and Firewall Builder's policy compiler translates
it to the model of the concrete firewall where it will be actually
deployed.

\subsection{Data Model}

We use an object model to represent various networking and security
concepts used in configuring firewalls. User data is saved in files
with \textit{.fwb} using syntax described in
section~\ref{syntax}. Objects are organized into {\sl Libraries}. Each
file is a collection of such libraries. Typically there is at least
one library of objects created by the user. Additionally, there is a
library of standard objects provided with Firewall Builder which
includes definitions of standard objects (such as a list of standard
address ranges for private networks per RFC 1918\cite{rfc1918}). When
used in a business environent, the company may supply some libraries
of company-wide objects to be used by all departements.

The objects could be rougly split into several categories:

\subsubsection{Basic Networking Objects}

This category includes some basic objects representing common concepts
used in Networking. Some of them are:

\begin{description}
\item[IPv4 Address] Internet Protocol (IP) version 4 address
\item[IP Service] IP service, defined by protocol number and some
  options like {\sl loose source rote} and {\sl record route}
\item[UDP Service] UDP service is defined by source and destination
  port ranges.
\item[TCP Service] TCP service is defined by source and destination
  port ranges and some flags.
\item[ICMP Service] ICMP service is defined by {\sl ICMP type} and
  {\sl ICMP code}
\item[Physical Address] Data link layer address, such as Ethhernet
  {\sl MAC address} or Frame Relay {\sl Data Link Connection
    Identifier (DLCI)}.
\item[Time Interval] Allows specify time period. Time intervals are
  commonly used to specify time-based firewall policies. It could be
  expressed either in terms of absolute date and time specifications,
  or in terms of week days (e.g. \textit{from Monday to Friday})
\end{description}

\subsubsection{Hosts, Firewalls, Policies}

More complex objects are {\sl Hosts}. They represent network
nodes (servers, workstations, routers, IP printers, etc.). Hosts can
have multiple {\sl interfaces} with static or dynamic IP addresses.

A {\sl Firewall} is a special kind of host, which will be running
firewall software and could be configured using Firewall Builder. The user
must specify what OS platform and firewall software they are using (some platforms
allow the user to select from several firewall packages). For firewalls, the
user can define a {\sl Firewall Policy} and {\sl NAT Rules}.

{\sl Firewall Policy} consists of a set of firewall {\sl rules}. Each
rule has a {\sl source} and {\sl destination}, {\sl service}, {\sl
  interface}, {\sl direction}, {\sl time} and an {\sl
  action}. Rule-matching semantics will be explained in
Section~\ref{packetprocessing}.

{\sl NAT Rules} specify how the firewall host performs network address
translation, changing sources and destinations of passing packets.

\subsubsection{Utility Objects}

Objects in these categories are various convenience objects,
representing higher-level concepts which are easy to use when
describing firewall policies.

\begin{description}
\item[Address Range] Range of IPv4 addesses. Specified by first and
  last address.
\item[Address Table] List to IPv4 addresses which is specified in an
  external file that can be loaded at policy compile time, or at the
  time of deployment of generated firewall policy, depending on the
  option configured in the Address Table object. Such lists are
  commonly used to maintain dynamically updated black lists of
  spammers or intruders.
\item[Groups] Various objects could be combined into named groups for the
  convenience of referencing them as such in policy rules. Users
  typically group hosts, IP addresses, services and time
  intervals. Groups are ``typed''. That means that groups can contain
  only objects of the same type.
\end{description}

\subsection{Syntax}
\label{syntax}

The policy is expressed as an Extensive Markup Language
(XML\cite{xml10}) document. The grammar of this document is specified
as a Document Type Definition (DTD) file. The DTD file for the current
version is shown in Appendix~\ref{dtd}.

Each object has an unique \emph{id} attribute. This attribute is used to
establish references between objects.

Here are some examples, to illustrate the syntax we use. First, some
simple objects:

\begin{lstlisting}[caption={\emph{Network} Object},frame=trBL]
<Network id="id47505CE816470" name="officeLAN" address="10.86.81.0" netmask="255.255.255.0"/>
\end{lstlisting}

\begin{lstlisting}[caption={\emph{UDP Service} Object},frame=trBL]
<UDPService id="id47505D0216470" name="MyServie" dst_range_end="92" dst_range_start="90" src_range_end="70" src_range_start="30"/>
\end{lstlisting}

Now let us take a look at a firewall with a simple single-rule
policy\footnote{for clarity, some non-essential
  attributes and elements were omitted}, shown on listing~\ref{fwex}.

\begin{lstlisting}[caption={\emph{Firewall} Object},frame=trBL,numbers=left, numberstyle=\tiny, label=fwex]
      <Firewall host_OS="linux24" id="id47505D0516470" name="MyFirewall" platform="iptables">
        <Interface dyn="False" id="id47505D0B16470" name="if0" unnum="False">
          <IPv4 address="192.168.1.1" id="id47505D0C16470" name="MyFirewall:if0:ip" netmask="255.255.255.0"/>
          <physAddress address="00:17:f2:ea:ee:35" id="id47505D3816470" name="MyFirewall:if0:mac"/>
        </Interface>
        <Interface dyn="True" id="id47505D0D16470" name="if1" unnum="False"/>
        <Interface dyn="False" id="id47505D0F16470" name="l0" unnum="False" unprotected="False">
          <IPv4 address="127.0.0.1" id="id47505D1016470" name="MyFirewall:l0:ip" netmask="255.255.0.0"/>
        </Interface>
        <Policy id="id47505D0816470">
          <PolicyRule action="Deny" comment="" direction="Both" disabled="False" id="id47505ECE16470" position="0">
            <Src neg="False">
              <ObjectRef ref="sysid0"/>
            </Src>
            <Dst neg="False">
              <ObjectRef ref="id47505CE816470"/>
            </Dst>
            <Srv neg="False">
              <ServiceRef ref="id47505D0216470"/>
            </Srv>
            <Itf neg="False">
              <ObjectRef ref="sysid0"/>
            </Itf>
            <When neg="False">
              <IntervalRef ref="sysid2"/>
            </When>
          </PolicyRule>
        </Policy>
      </Firewall>
\end{lstlisting}

As we can see, the \emph{Firewall} element includes the definition for
three network interfaces and a firewall policy. 

Interface definitions are expressed as \emph{Interface}
elements. Interface \emph{if1} is {\sl dynamic} and has no static IP
address associated with it. Interfaces \emph{if0} and \emph{lo0} have
static IP addresses assocated with them. These IP addresses are
expressed as enclosing \emph{IPv4} elements. One may wonder why
interface address was not specified as an attribute. The answer
is that an interface could have more than one IP address assigned to
it.

The firewall policy is expressed as a \emph{Policy} element, and may
contain one or more \emph{PolicyRule} elements. Because XML
specification \cite{xml10} does not guarantee element order, policy
rule ordering is implicitly specified via \emph{position} attrbute
which defines \emph{PolicyRule} absolute order within enclosing
\emph{Policy} element.

{\sl Direction} and {\sl Action} rule fields are specified via
\emph{direction} and \emph{action} attributes of a
\emph{PolicyRule}. Each \emph{PolicyRule} rule element contains
\emph{Src}, \emph{Dst}, \emph{Srv}, \emph{Itf}, \emph{When}
sub-elements to specify {\sl Source}, {\sl Destination}, {\sl
  Service}, {\sl Interface} and {\sl Time Interval} rule fields
respectively. Each of these elements could contain one or more object
references specifing their value.

Each of the field's matching value could optionally be made negative by
specifying \emph{neg} attribute. For example listing~\ref{noneg}
demonstrates a destination which is either an object with \emph{id}
\emph{A} or \emph{B}. Adding negation as shown on listing~\ref{yesneg}
changes the meaning so that the destination must be neither \emph{A}
nor \emph{B}.

\begin{lstlisting}[frame=trBL, label=noneg, caption={Negation Example (withou negation)}]
  <Dst neg="False">
      <ObjectRef ref="A"/>
      <ObjectRef ref="B"/>
  </Dst>
\end{lstlisting}

\begin{lstlisting}[frame=trBL, label=yesneg, caption={Negation Example (with negation)}]
  <Dst neg="True">
      <ObjectRef ref="A"/>
      <ObjectRef ref="B"/>
  </Dst>
\end{lstlisting}

As we have seen, there are two major ways to express relationships
between objects in the Firewall Builder XML. The first way is
embedding - when one object definition is enclosed in the other object
definition element. An example is an \emph{Interface} embedded within a
\emph{Firewall} object, or a \emph{IPv4} object, embedded within an
\emph{Interface} object. The second method uses a reference, via the
\emph{ObjectRef} element. In this method, in place of the object which
we are refering to, we place an \emph{ObjectRef} element, which has
its \emph{ref} attribute set to the value of \emph{id} of the object we
are reffering to. We can see such references in \emph{Src} and
\emph{Dst} elements of a \emph{PolicyRule} referencing \emph{Network}
and \emph{UDPService} objects respectively in the example above.

\subsection{Processing Model}
\label{packetprocessing}

It is not sufficient to define just {\sl Data Model} to be able to
write a firewall policy. A data model implies certain semantics,
defined as a {\sl processing model}. Processing model differs from one
firewall platform to another. We will define an abstract processing
model to be used when defining policies of Abstract Firewall and later
on we will map it to processing models of concrete firewall platforms.

For each packet passing through a firewall, several processing stages
are applied. It is optionally processed via NAT Rules and then
filtered by Firewall Policy Rules. These stages can change the packet
headers or even drop or reject the whole packet.

While the sequence of NAT and
filtering steps varies from platform to platform in real firewalls (see section
\ref{natfilteringorder} for disuccion), in Firewall Builder's abstract
firewall model, it is fixed and processing is always done in the
following order:

\begin{enumerate}
 \item Network Address Translation step is performed
 \item Firewall Policy is applied
\end{enumerate}

The packet is first matched towards all NAT rules, in the order they are
defined by the user. A NAT rule ``matches'' if the rule {\sl original
  source}, {\sl original destination} and {\sl original service}
fields match the current packet and if it happens within an optional time
interval specified in the rule. (any matching fields may be
specified as \emph{Any} - a wildcard which matches any value). A
matched packet is modified by replacing its {\sl source}, {\sl
  destination} and {\sl service} fields with {\sl translated source},
{\sl translated destination} and {\sl translated service} from the
rule. If some of {\sl translated source}, {\sl translated destination}
or {\sl translated service} is left empty by the user, it means that the
original value of this field should be preseved. If a packet has not
matched any NAT rules, it will be processed further, unchanged.

Next, the packet is matched towards all Policy rules in the order they
are defined by user. For each packet the following fields are matched
towards the rules:

\begin{description}
\item[Source] packet source address (IP or data link level)
\item[Destination] packet destination address (IP or data link level)
\item[Service] packet service (One of IP, UDP, ICMP service objects.)
\item[Interface] interface via which this packet has arrived
\item[Direction] direction of the packet, in respect to the firewall (\textit{Inbound} or \textit{Outbound})
\end{description}

Any of these field could be excluded from matching if \emph{Any}
wildcard is specified as the value in the rule.

Once a packet has matched one of the rules, the {\sl action} specified
in the rule is performed. Possible values are:

\begin{description}
\item[Accept] the packet is permitted to pass through
\item[Deny/Drop] the packet is silently dropped
\item[Reject] the packet is rejected, notyfing the server via ICMP message
\item[Accounting] the packet counter associated with this rule is incremented
\end{description}

Although actual firewall implementations may vary in what happens once a
packet is matched (see section~\ref{firstvslast} for examples), in the
Firewall Builder's abstract firewall model semantics are well defined:

\begin{quote}
For {\sl accept}, {\sl deny} and {\sl reject} actions after the first rule
is matched, the approriate action is performed and no further rule
checks are performed. For {\sl accounting} actions, after a counter
value increase, the packet matching is continued against any remaining
rules.
\end{quote}

After all rules have been processed and no {\sl accept}, {\sl deny} or
{\sl reject} action was invoked, the {\sl default policy} is applied.
While the default policy could be
different in underlying firewall platforms (see section~\ref{defaultpolicy} for discussion), in
Firewall Builder's abstract firewall model, the default policy is to perform
a {\sl drop} action on every packet.

\subsection{Policy Verification and Optimization}
\label{optandver}

Even before the policy is compiled to concrete firewall syntax, there
is certain processing which could be done on the abstract policy
model. The two main areas are \textit{verification} and
\textit{optimization}. Having well-defined processing model and a
policy expressed in a stadartized form, a generic high-level policy
analysis could be performed without needing to focus on the details of
firewall platform implementation.

\subsubsection{Verification}

While the XML syntax validation towards the DTD ensured that there are
no syntax errors in the document, it does not catch errors in semantics.

For example, we found it useful to show users a warning when some
policy rules will never be used. It is similar to unreachable code detection
in programming languages. For example, let us assume there are two
identical rules (with {\sl drop} action), which differ only in the
{\sl destination address} field. The first rule has destiation address
\textit{1.2.3.4/16} while the second rule has
\textit{1.2.3.4/32}. Obviously, all packets which could possibly match
the second rule, will be matched by the first rule first. We call this
situation ``rule shadowing'', saying that the first rule ``shadows''
the second one. We try to detect such situations and report them to the user,
since they most likely signify an error in the user's policy definition.

In addition to rule shadowing, in the future we can forsee other
semantic errors which can be detected and reported to the user. This is one of
the areas for the future research.

\subsubsection{Optimization}

There is a cost for executing each rule in a firewall. Long policies
tend to affect firewall performance. It is very beneficial to try to
optimize firewall policy by combining and reshuffling rules to make it
shorter and hence more efficient.

Common optimization techniques include removing unusued or redundant
rules, grouping multiple rules into a single one, and in general to try
to express the same policy with the fewer rules.

\section{Platform-specific challenges}
\label{challanges}

Let us examine selected examples of platform specifics on
{\sl pf}, {\sl iptables} and {\sl ipfilter} firewall platforms. All
these problems are normally hidden from Firewall Builder users, because
the firewall hides all these platform-specific differences from the user and
generates platform-specific code to resolve these issues.

\subsection{Implicit vs. Explicit Interface Specification}

\subsection{Default Policy}
\label{defaultpolicy}

What should a firewall do with a packet which matched none of the policy
rules? Should it be allowed to pass through, or should it be discarded?

In {\sl iptables} default policy is a user-configurable option.

In {\sl ipfilter} packets are also passed by default, unless it is
compiled with \textit{IPFILTER\_DEFAULT\_BLOCK}
option\cite{conoboy1911ifb}.

In {\sl pf} packets are passed by default

\subsection{First vs. Last Policy Rule Matching}
\label{firstvslast}

In typical packet filter, a packet is matched towards a list of rules. It
could either match or not match each rule. If a rule is matched, it makes a
decision to permit this packet ({\sl accept}) or not ({\sl
  reject} or {\sl drop}). There are two common matching
strategies. In the first strategy, matching occurs until the first matching
rule is found. We will call it {\sl first match}\footnote{This
  strategy is also sometimes reffered to as the ``single-trigger''
  approach}. Another strategy is to match all rules, then make a
decision based on \emph{last} match. We will call this strategy {\sl
  last match}\footnote{This strategy is also sometimes reffered to as
  ``mutli-trigger'' approach}.

{\sl iptables} supports only the {\sl first match} strategy.

{\sl ipfilter} and {\sl pf} both support {\sl last match} strategy by
default, unless \textit{quick} rule keyword was specified. This
keyword intstructs the firewall to stop further matching and use
results from the current match as a final decision on whenever packet should
be permitted to pass.

\subsection{NAT vs Firewall Rules Order}
\label{natfilteringorder}

Often a firewall will perform both \textit{packet filtering} and
\textit{network address translation (NAT)} functions. The obvious
question is: in what order NAT and filtering rules are applied? Are
addresses translated first and then filters are checked, or
vice-versa? This makes a big difference, because if NAT is applied first,
one should use already translated (not original) addresses in policy
rules.

{\sl iptables} destinguish two kind of NAT rules: SNAT (source NAT)
and DNAT (destination NAT). It could be said that DNAT is applied
first, then packet filtering, and then SNAT.

{\sl PIX}, another popular firewall platform from CISCO performs
packet filtering first and then NAT.

Both {\sl ipfilter} and {\sl pf} perform address translation first and
only then perform filtering functions.\footnote{In case of {\sl pf}:
  ``The only exception to this rule is when the \textit{pass} keyword
  is used within the nat rule. This will cause the NATed packets to
  pass right through the filtering engine.''\cite{pffaq}}

\subsection{Negation}

Sometimes it is convenient to use negation in policy rules. For
example, to specify condition like ``if source address is
\emph{not} {\sl 1.2.3.4}''. A more complex form of negation is to apply it to
a group of addresses (``if source address is not in \{{\sl
  1.2.3.4}, {\sl 10.20.30.40}\}'').

{\sl iptables} support single address negation: 

\begin{quote}
  ``Many flags, including the \textit{`-s'} (or \textit{`--source'})
  and \textit{`-d'} (\textit{`--destination'}) flags can have their
  arguments preceded by \textit{`!'} (pronounced \textit{`not'}) to
  match addresses \emph{NOT} equal to the ones given. For
  example. \textit{`-s ! localhost'} matches any packet not coming
  from \textit{localhost}.''\cite{russell2004lpf}.
\end{quote}

However for address ranges, support for which is facilitated by {\sl
  mod\_iprange} module, negation is not suported.

Both {\sl ipfilter} supports negation (at least for addresses). No
group negation support is provided.

{\sl pf} supports negation (at least for addresses). It also supports a
limited case of group negation, when using tables. For example, the
following fragment allows to pass all trafic from all addresses,
except ones in the black list.

\begin{small}
\begin{verbatim}
table <blacklist> {1.2.3.4 , 10.20.30.40}
pass in quick inet from any to ! <blacklist> keep state
\end{verbatim}
\end{small}

\subsection{Addrress Range Emulation}

All firewalls allow the user to specify an individual IP address or {\sl CIDR
  block} in the rules. However, sometimes it is convenient to specify an
address range (\textit{from - to}).

{\sl iptables} permits address ranges using {\sl iprange} module.

Both {\sl ipfilter} and {\sl pf} do not support address ranges.

\subsection{Dymanic Interfaces}

Oftentimes, the IP address assigned to an interface is not known at the
time of the policy definition. This is common with {\sl dynamic
  interface}, which obtains its address using DHCP or a similar
protocol. Abstract Firewall Policy allows the user to implement such intefaces in
policy rules, in place of source or destination addresses.

{\sl pf} permits the use of inteface names in the rules, and will use
current interface IP addresses at the time the rule is executed.

{\sl ipfilter} is using special \textit{0/32} notation to refer to
currently assigned interface IP address.

In the case of {\sl iptables} there is no way to refer to the current
interface dynamically-assigned IP address in policy rules.

\section{Abstract Policy Compilation Techniques}
\label{processingtechniques}

In this section we will briefly discuss some implementation approaches
used to compile and deploy Abstract Firewall policy to a concrete
firewall platform.

An abstract firewall policy needs to be compiled into policy for the
concrete firewall. Usually this requires certain
transformations. While overall rule data structure remains rougly the
same (source, destitatio, action, etc.), a target firewall platform
puts various limitations to the allowed values, and sometimes even implies
slightly different semantics.

We found it convenient to perform policy transformation as a series of
small steps. Each step could be viewed as a function, which takes as
input a list of policy rules and outputs a modified list of such
rules. Some of these transformations are quite simple and could be
reused between different firewall platforms. These transformation
functions are called {\sl Rule Processors}. An example of a rule
processor could be one which takes a single rule with {\sl address
  ranges} in the rule {\sl source address} and converts it to a group
of rules, which together perform the same function as the original rule, but
each rule has a single {\sl CIDR} block in a {\sl source address
  field}.

\section{Related Work}

There is a lot of related research in this area (see
\cite{elatawy60604suf} for a good survey on the subject).

Many approaches are concentrated on building an abstract security model,
and then applying to to the firewall policies (either automated
genetation or verification). Some models are using UML, some build
upon RBAC model.

In our opinion, one of the problems with such approaches is the big
representation gap between the model abstractions and the concrete
firewall device processing and data model. Our approach is more
pragmatic. Firewall Builder's abstract firewall model is very close to
the one used in the many modern day firewall devices. This model is
familiar to the most firewall administrators and easy to
understand. Our model could act as intermediate representation
between high-level models and formal languages and concrete firewall
policies.

Al-Shaer et. al\cite{alshaer2004dpa} present good formalization of
firewall rules relationships and classification of the {\sl anomalies}
which should be detected during policy verification.

\section{Conclusions}
\label{future}

In this paper we have presented in an overview form Firewall Builder's
approach of corss-platform firewall management: the idea of an
Abstract Firewall, the data and a processing model of such
firewall. In a few examples we have shown the kind of challenges firewall
adminstrators are facing when they are required to work with multiple firewall
platforms.

The definition of an abstract firewall model and policy definition language
is a first, enabling step which allows us to develop and apply various
policy analysis and transformation techniques in a
platform-independent manner. Policiy verification and optimization
techniques, briefly touced upon in the section~\ref{optandver} presents
many interesting research challenges and opportunities.

Firwall Builder data files could contain multiple firewalls sharing
common utility objects (hosts, networks, etc.). This opens the
opportunity for developing more sophisticated policy analysis tools,
considering not only a single firewall but a network with several
firewalls. Such a comprehensive distributed firewal model could be
analyzed for {\sl inter-firewall anomalies} as well as {\sl
  intra-firewall anomalies}\cite{alshaer2004dpa}.

In the course of the project, we started to work on a formal model of
policy rules relationships. Such a model is required to implement
non-trivial validation and optimization techniques. Our initial
thinking was along the lines of multi-dimensional space, where each
rule field represents a dimension and a rule represents a figure. Each
packet is represented as a point in this space. If it matches some rule,
this point will be inside a figure represented by the rule.

\nocite{*}
\bibliographystyle{plain}
\bibliography{FWBPolicy}

\begin{thebibliography}{10}

\bibitem{fwb}
Firewall builder project.
\newblock http://www.fwbuilder.org/.

\bibitem{pffaq}
Pf: The openbsd packet filter.
\newblock http://www.openbsd.org/faq/pf/.

\bibitem{alshaer2004dpa}
E.~Al-Shaer and H.~Hamed.
\newblock {Discovery of policy anomalies in distributed firewalls}.
\newblock {\em IEEE INFOCOM}, 4:2605--2616, 2004.

\bibitem{bauer2003ppu1}
M.~Bauer.
\newblock {Paranoid penguin: using firewall builder, Part I}.
\newblock {\em Linux Journal}, 2003(109), 2003.

\bibitem{bauer2003ppu2}
M.~Bauer.
\newblock {Paranoid Penguin: Using Firewall Builder, Part II}.
\newblock {\em Linux Journal}, 2003(110), 2003.

\bibitem{xml10}
T.~Bray, J.~Paoli, C.M. Sperberg-McQueen, et~al.
\newblock {Extensible Markup Language (XML) 1.0}.
\newblock {\em W3C Recommendation}, 6, 2000.

\bibitem{conoboy1911ifb}
B.~Conoboy and E.~Fichtner.
\newblock {IP Filter Based Firewalls HOWTO}.
\newblock {\em Sat}, 22(26):2001, 1911.

\bibitem{cuppens2005dar}
F.~Cuppens, N.~Cuppens-Boulahia, and J.~Garc{\i}a-Alfaro.
\newblock {Detection and Removal of Firewall Misconfiguration}.
\newblock {\em Proceedings of the 2005 IASTED International Conference on
  Communication, Network and Information Security (CNIS 2005)}, 2005.

\bibitem{elatawy60604suf}
A.~El-Atawy.
\newblock {Survey on the use of formal languages/models for the specification,
  verification, and enforcement of network access-lists}.
\newblock {\em School of Computer Science, Telecommunication, and Information
  Systems, DePaul University, Chicago, Illinois}, 60604.

\bibitem{kurland2003fb}
V.~Kurland.
\newblock {Firewall Builder}.
\newblock {\em 11th DFN-CERT Workshop, Hamburg, Germany}, 2004.

\bibitem{fbug2003}
NetCitadel LLC.
\newblock {Firewall Builder User's Guide}.
\newblock 2003.

\bibitem{rfc1918}
Y.~Rekhter, B.~Moskowitz, D.~Karrenberg, GJ~de~Groot, and E.~Lear.
\newblock {RFC1918: Address Allocation for Private Internets}.
\newblock {\em Internet RFCs}, 1996.

\bibitem{russell2004lpf}
R.~RUSSELL.
\newblock Linux 2.4 packet filtering howto, 2004.

\bibitem{fwbupgrades}
V.~Zaliva.
\newblock Managing xml documents versions and upgrades with xslt.
\newblock 2001.

\end{thebibliography}

\singlespacing

\appendix
\appendixpage
\addappheadtotoc

\section{Firewall Builder DTD}
\label{dtd}

\begin{tiny}
\begin{verbatim}
<?xml version="1.0" encoding="utf-8"?>
<!--
     Firewall Builder Document Type Definition
     http://www.fwbuilder.org/
     Version: $Revision: 1.41 $
     Authors:  Friedhelm Duesterhoeft, Vadim Zaliva, Vadim Kurland, Tidei Maurizio

TODO:

1. Allow groups of unrelated objects.

-->

<!ENTITY % BOOLEAN   "(False|True)">
<!ENTITY % STRING    "CDATA">
<!ENTITY % NUMBER    "CDATA">

<!--
 * Supported policy rule actions:
 *
 *        Accept - accept the packet, analysis terminates
 *
 *        Reject - reject the packet and send ICMP 'unreachable' or
 *                 TCP RST back to sender, analysis terminates
 *
 *        Deny   - drop the packet, nothing is sent back to sender,
 *                 analysis terminates
 *
 *        Scrub  - run the packet through normalizer (see 'scrub' in
 *                 PF), continue analysis
 *
 *        Return - action used internally, meaning may depend on
 *                 implementation of the policy compiler but generally 
 *                 means return from the block of rules
 *
 *        Skip   - skip N rules down and continue analysis. Used
 *                 internally.
 *
 *        Continue - do nothing, continue analysis. Used internally.
 *
 *        Accounting - generate target firewall platform rule to count
 *                 the packet and continue analysis.
 *
 *        Modify - edit the packet (change some header values, like
 *                 TOS bits) or mark it somehow if the kernel supports 
 *                 that (e.g. target MARK in iptables)
 *
 *        Tag    - put a tag on the packet or mark it somehow
 *        
 *        Pipe   - send the packet to the userland process for inspection
 *        
 *        Classify - classify the packet for QoS or traffic shaping
 * 
 *        Custom - platform-depended custom action
 *
 *        Branch - branch to a subset of rules for inspection
 *
-->

<!ENTITY % ACTION    "(Accept|Reject|Deny|Scrub|Return|Skip|Continue|Accounting|Modify|Tag|Pipe|Classify|Custom|Branch|Route)">
<!ENTITY % DIRECTION "(Inbound|Outbound|Both)">
<!ENTITY % IPADDRESS "CDATA">
<!ENTITY % NETMASK   "CDATA">

<!-- Standard attributes presented in all nodes -->
<!ENTITY % STD_ATTRIBUTES '
 name    %STRING;  #REQUIRED
 comment %STRING;  #IMPLIED
 id      ID        #REQUIRED
 ro      %BOOLEAN; #IMPLIED
'>

<!-- Standard attributes for all system nodes -->
<!ENTITY % SYS_ATTRIBUTES '
'>

<!-- 
      **** Document structure, main groups. ****
-->

<!ELEMENT FWObjectDatabase (Library*)>
<!ATTLIST FWObjectDatabase
 xmlns        CDATA     #FIXED "http://www.fwbuilder.org/1.0/"
 version      %STRING;  #FIXED "2.1.14"
 lastModified %NUMBER;  #IMPLIED
 id           ID        #REQUIRED
>

<!ELEMENT Library ((AnyNetwork|AnyIPService|AnyInterval|ObjectGroup|Host|Firewall|
Network|IPv4|DNSName|AddressTable|physAddress|AddressRange|ObjectRef|ServiceGroup|
IPService|ICMPService|TCPService|UDPService|CustomService|ServiceRef|
IntervalGroup|Interval|IntervalRef|Interface|Policy|NAT|PolicyRule|
NATRule|Library|TagService)*)>
<!ATTLIST Library
 %STD_ATTRIBUTES;
 color   %STRING;  #IMPLIED
>


<!-- 
      **** Document structure, Services. ****
-->

<!ELEMENT AnyIPService EMPTY>
<!ATTLIST AnyIPService
 %SYS_ATTRIBUTES;
 %STD_ATTRIBUTES;
 protocol_num %NUMBER;  #FIXED     "0"
>

<!-- Reference to Services child -->
<!ELEMENT ServiceRef EMPTY>
<!ATTLIST ServiceRef 
       ref IDREF #REQUIRED
>

<!ELEMENT ServiceGroup (( ServiceGroup | IPService | ICMPService  | TCPService | UDPService | CustomService | ServiceRef | TagService)*)>
<!ATTLIST ServiceGroup
 %STD_ATTRIBUTES;
>

<!-- 
      **** Document structure, Objects. ****
-->

<!-- Reference to Objects child -->
<!ELEMENT ObjectRef EMPTY>
<!ATTLIST ObjectRef 
       ref IDREF #REQUIRED
>

<!ELEMENT ObjectGroup ((ObjectGroup|Host|Firewall|Network|IPv4|DNSName|AddressTable|AddressRange|ObjectRef)*)>
<!ATTLIST ObjectGroup
 %STD_ATTRIBUTES;
>

<!-- 

This element will contain elements with platform  specific
options.

<Options>
    <Option name="option1_name">Value1</Option>
    <Option name="option2_name">Value2</Option>
</Options>

Since list of compilers is open (everybody could write his
own compiler) we do not define content model for this element.

-->

<!ELEMENT Option ANY>
<!ATTLIST Option
 name %STRING; #REQUIRED
>


<!ELEMENT PolicyRuleOptions (Option*)>
<!ELEMENT NATRuleOptions    (Option*)>
<!ELEMENT RoutingRuleOptions (Option*)>
<!ELEMENT FirewallOptions   (Option*)>
<!ELEMENT HostOptions       (Option*)>
<!ELEMENT GatewayOptions    (Option*)>

<!-- 
      **** Document structure, rest ****
-->

<!ELEMENT NATRule (OSrc,ODst,OSrv,TSrc,TDst,TSrv,When?, NATRuleOptions?, NAT?)>
<!ATTLIST NATRule
 id      ID       #REQUIRED
 disabled  %BOOLEAN;   "False"
 position  %NUMBER;    #REQUIRED
 comment   %STRING;    #IMPLIED
>

<!ELEMENT When (IntervalRef*)>
<!ATTLIST When
 neg %BOOLEAN; #REQUIRED
>

<!ELEMENT OSrc (ObjectRef*)>
<!ATTLIST OSrc
 neg %BOOLEAN; #REQUIRED
>

<!ELEMENT ODst (ObjectRef*)>
<!ATTLIST ODst
 neg %BOOLEAN; #REQUIRED
>

<!ELEMENT OSrv (ServiceRef*)>
<!ATTLIST OSrv
 neg %BOOLEAN; #REQUIRED
>

<!ELEMENT TSrc (ObjectRef*)>
<!ATTLIST TSrc
 neg %BOOLEAN; #REQUIRED
>

<!ELEMENT TDst (ObjectRef*)>
<!ATTLIST TDst
 neg %BOOLEAN; #REQUIRED
>

<!ELEMENT TSrv (ServiceRef*)>
<!ATTLIST TSrv
 neg %BOOLEAN; #REQUIRED
>


<!ELEMENT RoutingRule (RDst,RGtw,RItf, RoutingRuleOptions?, Routing?)>
<!ATTLIST RoutingRule
 id      ID       #REQUIRED
 disabled  %BOOLEAN;   "False"
 position  %NUMBER;    #REQUIRED
 metric    %NUMBER;    "0"
 comment   %STRING;    #IMPLIED
>

<!ELEMENT RDst (ObjectRef*)>
<!ATTLIST RDst
 neg %BOOLEAN; #REQUIRED
>

<!ELEMENT RGtw (ObjectRef*)>
<!ATTLIST RGtw
 neg %BOOLEAN; #REQUIRED
>

<!ELEMENT RItf (ObjectRef*)>
<!ATTLIST RItf
 neg %BOOLEAN; #REQUIRED
>


<!ELEMENT PolicyRule (Src,Dst,Srv?,Itf?,When?,PolicyRuleOptions?,Policy?)>
<!ATTLIST PolicyRule
 id       ID       #REQUIRED
 disabled  %BOOLEAN;   "False"
 position  %NUMBER;    #REQUIRED
 direction %DIRECTION; #IMPLIED
 action    %ACTION;    #REQUIRED
 log       %BOOLEAN;   #REQUIRED
 comment   %STRING;    #IMPLIED
>

<!ELEMENT Src (ObjectRef*)>
<!ATTLIST Src
 neg %BOOLEAN; #REQUIRED
>

<!ELEMENT Dst (ObjectRef*)>
<!ATTLIST Dst
 neg %BOOLEAN; #REQUIRED
>

<!ELEMENT Srv (ServiceRef*)>
<!ATTLIST Srv
 neg %BOOLEAN; #REQUIRED
>

<!ELEMENT Itf (ObjectRef*)>
<!ATTLIST Itf
 neg %BOOLEAN; #REQUIRED
>


<!--
   hardware or physical address (MAC, DLCI etc.)
-->
<!ELEMENT physAddress EMPTY>
<!ATTLIST physAddress
 %STD_ATTRIBUTES;
 address %STRING; #REQUIRED
>

<!ELEMENT IPv4 EMPTY>
<!ATTLIST IPv4
 %STD_ATTRIBUTES;
 address %IPADDRESS; #REQUIRED
 netmask %NETMASK;   #REQUIRED
>

<!ELEMENT DNSName EMPTY>
<!ATTLIST DNSName
 %STD_ATTRIBUTES;
 dnsrec         %STRING;    #REQUIRED
 run_time       %BOOLEAN;   #REQUIRED
>

<!ELEMENT AddressTable ((IPv4|ObjectRef)*)>
<!ATTLIST AddressTable
 %STD_ATTRIBUTES;
 filename       %STRING;    #REQUIRED
 run_time       %BOOLEAN;   #REQUIRED
>
<!--
Interface can have the following attributes:

   - dyn            interface has dynamically assigned address
   - unnum          interface is unnumbered (does not have IP address, but 
                    may still have MAC address)
   - bridgeport     interface serves as a bridge port on bridging firewall.
                    The difference between bridge port and unnumbered interface
                    is that compilers may use special modules or commands for
                    bridge ports on platforms that support them, such as 
                    module physdev for iptables.
   - mgmt           this is management interface
   - physAddress    MAC address of this interface
   - security_level
   - network_zone   ID of the object representing network zone
   - unprotected    Skip this interface while assigning access lists or policy rules
   - label          human-readable label of this interface

-->
<!ELEMENT Interface (IPv4*, physAddress?)>
<!ATTLIST Interface
 %STD_ATTRIBUTES;
 dyn            %BOOLEAN;   #REQUIRED
 unnum          %BOOLEAN;   #IMPLIED
 mgmt           %BOOLEAN;   #IMPLIED
 bridgeport     %BOOLEAN;   #IMPLIED
 security_level %NUMBER;    #REQUIRED
 network_zone   IDREF       #IMPLIED
 unprotected    %BOOLEAN;   #IMPLIED
 label          %STRING;    #IMPLIED
>


<!-- Remote management information for Firewall, Host, Gateway  -->
<!ELEMENT Management (SNMPManagement? , FWBDManagement?, PolicyInstallScript?)>
<!ATTLIST Management
 address              %IPADDRESS; #REQUIRED
>

<!-- User-defined custom policy installation script for Firewall  -->
<!ELEMENT PolicyInstallScript EMPTY>
<!ATTLIST PolicyInstallScript
 enabled   %BOOLEAN;  "False"
 command   %STRING;    #IMPLIED
 arguments %STRING;    #IMPLIED
>

<!-- SNMP management information for Firewall, Host, Gateway  -->
<!ELEMENT SNMPManagement EMPTY>
<!ATTLIST SNMPManagement
 enabled  %BOOLEAN;  "False"
 snmp_read_community  %STRING;    #IMPLIED
 snmp_write_community %STRING;    #IMPLIED
>

<!-- FWBD management information for Firewall, Host, Gateway  -->
<!ELEMENT FWBDManagement (PublicKey?)>
<!ATTLIST FWBDManagement
 enabled  %BOOLEAN;  "False"
 port                 %NUMBER;    #REQUIRED
 identity             %STRING;    #REQUIRED
>

<!-- Remote FWBD public key for Firewall, Host, Gateway  -->
<!ELEMENT PublicKey (#PCDATA)>

<!ELEMENT Host (Interface*, Management?, HostOptions?)>
<!ATTLIST Host
 %STD_ATTRIBUTES;
 host_OS              %STRING;    #IMPLIED
>

<!ELEMENT AnyNetwork EMPTY>
<!ATTLIST AnyNetwork
 %SYS_ATTRIBUTES;
 %STD_ATTRIBUTES;
 address              %IPADDRESS; #FIXED    "0.0.0.0"
 netmask              %NETMASK;   #FIXED    "0.0.0.0"
>

<!ELEMENT Network EMPTY>
<!ATTLIST Network
 %STD_ATTRIBUTES;
 address %IPADDRESS; #REQUIRED
 netmask %NETMASK;   #REQUIRED
>

<!ELEMENT AddressRange EMPTY>
<!ATTLIST AddressRange
 %STD_ATTRIBUTES;
 start_address %IPADDRESS; #REQUIRED
 end_address   %IPADDRESS; #REQUIRED
>

<!ELEMENT ICMPService EMPTY>
<!ATTLIST ICMPService
 %STD_ATTRIBUTES;
 code %NUMBER; #REQUIRED
 type %NUMBER; #REQUIRED
>


<!ELEMENT TagService EMPTY>
<!ATTLIST TagService
 %STD_ATTRIBUTES;
 tagcode %STRING;  #REQUIRED
>

<!ELEMENT IPService EMPTY>
<!ATTLIST IPService
 %STD_ATTRIBUTES;
 protocol_num %NUMBER;  #REQUIRED
 fragm        %BOOLEAN; #IMPLIED
 lsrr         %BOOLEAN; #IMPLIED
 rr           %BOOLEAN; #IMPLIED
 short_fragm  %BOOLEAN; #IMPLIED
 ssrr         %BOOLEAN; #IMPLIED
 ts           %BOOLEAN; #IMPLIED
>

<!ELEMENT TCPService EMPTY>
<!ATTLIST TCPService
 %STD_ATTRIBUTES;
 dst_range_end   %NUMBER;  #REQUIRED
 dst_range_start %NUMBER;  #REQUIRED
 urg_flag        %BOOLEAN; #REQUIRED
 ack_flag        %BOOLEAN; #REQUIRED
 psh_flag        %BOOLEAN; #REQUIRED
 rst_flag        %BOOLEAN; #REQUIRED
 syn_flag        %BOOLEAN; #REQUIRED
 fin_flag        %BOOLEAN; #REQUIRED
 urg_flag_mask   %BOOLEAN; #REQUIRED
 ack_flag_mask   %BOOLEAN; #REQUIRED
 psh_flag_mask   %BOOLEAN; #REQUIRED
 rst_flag_mask   %BOOLEAN; #REQUIRED
 syn_flag_mask   %BOOLEAN; #REQUIRED
 fin_flag_mask   %BOOLEAN; #REQUIRED
 src_range_end   %NUMBER;  #REQUIRED
 src_range_start %NUMBER;  #REQUIRED
 established     %BOOLEAN; #IMPLIED
>

<!ELEMENT UDPService EMPTY>
<!ATTLIST UDPService
 %STD_ATTRIBUTES;
 dst_range_end   %NUMBER; #REQUIRED
 dst_range_start %NUMBER; #REQUIRED
 src_range_end   %NUMBER; #REQUIRED
 src_range_start %NUMBER; #REQUIRED
>

<!ELEMENT CustomServiceCommand (#PCDATA)>
<!ATTLIST CustomServiceCommand
 platform %STRING; #REQUIRED
>

<!ELEMENT CustomService  (CustomServiceCommand*)>
<!ATTLIST CustomService
 %STD_ATTRIBUTES;
>

<!ELEMENT Gateway (Interface* , Management?, GatewayOptions?)>
<!ATTLIST Gateway
 %STD_ATTRIBUTES;
 address              %IPADDRESS; #REQUIRED
 host_OS              %STRING;    #IMPLIED
>

<!ELEMENT Firewall (NAT , Policy , Routing , Interface* , Management?, FirewallOptions?)>
<!ATTLIST Firewall
 %STD_ATTRIBUTES;
 platform             %STRING;    #REQUIRED
 version              %STRING;    #IMPLIED
 host_OS              %STRING;    #IMPLIED
 lastModified         %NUMBER;    #IMPLIED
 lastInstalled        %NUMBER;    #IMPLIED
 lastCompiled         %NUMBER;    #IMPLIED
 inactive             %BOOLEAN;   #IMPLIED
>

<!ELEMENT NAT (NATRule*)>
<!ATTLIST NAT
 id       ID       #REQUIRED
>

<!ELEMENT Policy (PolicyRule*)>
<!ATTLIST Policy
 id       ID       #REQUIRED
>

<!ELEMENT Routing (RoutingRule*)>
<!ATTLIST Routing
 id       ID       #REQUIRED
>


<!-- Time -->

<!ELEMENT IntervalGroup ((IntervalGroup|Interval|IntervalRef)*)>
<!ATTLIST IntervalGroup
 %STD_ATTRIBUTES;
>

<!-- Reference to time interval -->
<!ELEMENT IntervalRef EMPTY>
<!ATTLIST IntervalRef 
       ref IDREF #REQUIRED
>

<!ELEMENT Interval EMPTY>
<!ATTLIST Interval
 %STD_ATTRIBUTES;
 from_second  %NUMBER; "-1"
 from_minute  %NUMBER; "-1"
 from_hour    %NUMBER; "-1"
 from_day     %NUMBER; "-1"
 from_month   %NUMBER; "-1"
 from_year    %NUMBER; "-1"
 from_weekday %NUMBER; "-1"

 to_second    %NUMBER; "-1"
 to_minute    %NUMBER; "-1"
 to_hour      %NUMBER; "-1"
 to_day       %NUMBER; "-1"
 to_month     %NUMBER; "-1"
 to_year      %NUMBER; "-1"
 to_weekday   %NUMBER; "-1"
>

<!ELEMENT AnyInterval EMPTY>
<!ATTLIST AnyInterval
 %SYS_ATTRIBUTES;
 %STD_ATTRIBUTES;
 from_second  %NUMBER; #FIXED "-1"
 from_minute  %NUMBER; #FIXED "-1"
 from_hour    %NUMBER; #FIXED "-1"
 from_day     %NUMBER; #FIXED "-1"
 from_month   %NUMBER; #FIXED "-1"
 from_year    %NUMBER; #FIXED "-1"
 from_weekday %NUMBER; #FIXED "-1"

 to_second    %NUMBER; #FIXED "-1"
 to_minute    %NUMBER; #FIXED "-1"
 to_hour      %NUMBER; #FIXED "-1"
 to_day       %NUMBER; #FIXED "-1"
 to_month     %NUMBER; #FIXED "-1"
 to_year      %NUMBER; #FIXED "-1"
 to_weekday   %NUMBER; #FIXED "-1"
>

\end{verbatim}
\end{tiny}

\end{document}